# Negative Refraction and Subwavelength Lensing in a Polaritonic Crystal

*X. Wang and K. Kempa*

*Department of Physics, Boston College*

*Chestnut Hill, MA 02467*

We show that a two-dimensional polaritonic crystal, made of metallic rods that support well defined plasmon oscillations, can act in a narrow frequency range as a medium in which a negative refraction and subwavelength lensing can occur. We show that surface modes are excited on the surface of the lens, and that they facilitate restoration of the evanescent waves, which carry the subwavelength image information. We demonstrate that this can occur in the visible frequency range, for a wide range of materials, including silver and aluminum rods, and carbon nanotubes. This flexibility should allow for an experimental demonstration of this phenomenon in the visible frequency range.

wangxb@bc.edu

October 21, 2004

The early study of left-handed materials (LHM) by Veselago [1] in the late 60's, has led recently to an upsurge of research activity after Pendry showed that the negative refraction in such materials can lead to superlensing [2-9]. Two kinds of artificial materials have been proposed to simulate the LHM. First, following the proposal by Pendry [3], is an effective medium made of conducting wires and loops, suspended in a dielectric matrix. While wires assure an effective negative dielectric constant, the various loops provide a negative magnetic permeability in the same frequency band [8,9]. The second kind of the artificial LHM is based on a two-dimensional photonic crystal (2DPC) [10-13], a medium with an in-plane periodic modulation of the dielectric constant, which under proper conditions can act as an effective medium with a homogeneous refractive index $n = -1$. In such a case, a superlensing can occur in a simple slab of the 2DPC, with an image construction that follows precisely the simple rules of geometric optics [13]. A restricted superlensing can also occur, but in this case the image formation is complex, and restricted to a close vicinity of the slab surface [14,15]. The restricted superlensing has been confirmed experimentally at microwave frequencies [16,17].

The basic problem in achieving the superlensing at visible frequencies lies in the stringent requirement of the sufficient dielectric modulation in the 2DPC. The required dielectric modulation amplitude is at least 9, impossible to fulfill in the visible range with the existing low loss materials. In this letter we propose to overcome this problem by employing a two-dimensional *polaritonic* crystal (2DPLC) made of metallic units, each supporting a plasma mode that interferes with photons, forming a plasmon polariton. As a result of this interaction, photons "acquire mass", and that modifies their dispersion leading to highly symmetric concave, and also convex polaritonic bands ideal for superlensing, at small dielectric modulation amplitudes. We show, that this leads to an unrestricted subwavelength lensing at visible frequencies, in a structure made of available materials.

The 2DPLC studied here is a lattice of infinitely long metallic rods, immersed in a background matrix with a dielectric constant $\varepsilon$. The radius of each rod is $r$. Electromagnetic response of each rod is described by a Drude dielectric function, given by

$$\varepsilon(\omega) = \varepsilon_b - \frac{\omega_p^2}{\omega(\omega + i\Gamma)} \tag{1}$$

where $\varepsilon_b$ is the background dielectric constant, and $\omega$, $\Gamma$ are the radiation, and scattering frequencies, respectively, and where the plasma frequency is

$$\omega_p = \sqrt{\frac{4\pi n e^2}{m}} \tag{2}$$

where $n$ is the electron density, $m$ is the electron mass, and $e$ is the electron charge. Eq.(1) represents the minimal, realistic account of the plasmonic nature of a metal. For $\Gamma = 0$, $\varepsilon(\omega_p) = 0$, i.e. longitudinal mode, the plasmon, exists in the system. This electronic excitation couples directly to the photon field, leading to formation of the plasmon polariton.

An estimate of this effect can be given by following the analysis of Pendry [3], or the dipolar analysis of Ref. [18]. An array of long wires can be viewed as an effective medium, which can still be described by the Eq. (1), but with $\varepsilon_b = \varepsilon$, and the plasma frequency given by Eq. (2) with $n$ replaced by an effective electron density $n_{eff} = pn$, where $p = S_r/S$ is the filling factor, the ratio of the total cross-sectional area of the rods $S_r$ to the total cross-sectional area $S$ of the 2DPLC. It can be shown, that the modification of the electron mass $m$ in Eq. (2), due to the inter-wire induction [3] is very small here, and can be neglected. With this modification, Eq. (1) can be inserted into the usual photon dispersion $\omega = kc/\sqrt{\varepsilon(\omega)}$ ($k$ is the photon momentum and $c$ is the speed of light) to yield the plasmon-polariton dispersion, which for $\Gamma = 0$ has a simple form

$$\omega = \sqrt{\omega_p^2/\varepsilon + k^2 c^2/\varepsilon} \tag{3}$$

When there are no plasmons in the system ($\omega_p = 0$), Eq. (3) yields an acoustic branch $\omega = kc$ (massless photon). For $\omega_p \neq 0$, there is a gap of the size $\omega_p/\sqrt{\varepsilon}$ in the spectrum, and the corresponding photonic dispersion is parabolic in the vicinity of the gap edge $\omega = \omega_p / \sqrt{\varepsilon} + k^2 c^2 / (2\sqrt{\varepsilon}\omega_p)$. Thus, the photon acquires a mass as it transforms itself, due to the interaction, into the coupled plasmon-polariton mode. The detail analysis below, shows that the plasmon-polariton bands are parabolically soften at *all* the band edges, and thus one obtains also highly symmetric convex equal-frequency surfaces, ideal for superlensing.

To study the band structure of both, the photonic and polaritonic crystals in detail, we use the finite-difference time-domian (FDTD) [19] method. We consider the so-called polarization equation approach. The form of the Drude dielectric function (Eq. 1) implies the following differential equation

$$\frac{\partial^2 \mathbf{P}}{\partial t^2} - \Gamma \frac{\partial \mathbf{P}}{\partial t} = \varepsilon_0 \omega_p^2 \mathbf{E} \qquad (4)$$

where $\mathbf{P} = \varepsilon_0 [\varepsilon(\omega) - 1] \mathbf{E}$ is the polarization. By introducing

$$\frac{\partial \mathbf{P}}{\partial t} = \mathbf{J} \qquad (5)$$

we obtain,

$$\frac{\partial \mathbf{J}}{\partial t} - \Gamma \mathbf{J} = \varepsilon_0 \omega_p^2 \mathbf{E} \qquad (6)$$

Note that all the coefficients in the above equations are frequency-independent. The equations (5) and (6) can be solved numerically together with Maxwell's equations,

$$\frac{\partial \mathbf{H}}{\partial t} = -\frac{1}{\mu_0} \nabla \times \mathbf{E} \qquad (7)$$

$$\frac{\partial \mathbf{E}}{\partial t} = \frac{1}{\varepsilon_0} [\nabla \times \mathbf{H} - \mathbf{J}] \qquad (8)$$

by discretizing space and time, and by replacing partial derivatives with the centered finite differences. The electric and magnetic fields are defined on two spatially interleaved grids. The dielectric function and the magnetic permeability are defined at the grid points. We set $\omega_p = 0$ when the point is in background dielectric material. The unit cell contains 6400 ($80 \times 80$) grid points, which guarantees good convergence. To obtain the band structure, we Fourier-transform the fields form the time to the frequency domain.

To study the lensing, we employ the perfectly matched, layer boundary conditions to efficiently absorb the outgoing waves. [20].

Fig. 1 shows the result of band structure calculations for a triangular lattice of metallic rods with $r = 0.3a$ (solid line, $a$ is the lattice constant), for the transverse magnetic (TM) modes, and for wave vectors along the edges of the irreducible Brillouin zone (B.Z.) defined by the points $\Gamma$, M and K. The dimensionless frequency is $\Omega = \omega a / 2\pi c$, and the plasma frequency is $\Omega_p = 1$, and $\Gamma = 0$. The inter-rod space is filled by a medium with a dielectric constant $\varepsilon = 5$. For comparison, we show the lowest branch for the crystal with $r = 0.25a$ (dashed line). While the lowest band for a typical 2D photonic crystal consists of straight photon lines (for example see Ref. 13), a large gap exists below the first band for any 2DPLC. The simplified dispersion given by Eq. (3), shown as circles in Fig. 1, is in excellent agreement with the exact result. The upper branch of the simplified dispersion is obtained by an Umklapp of the lower branch outside the B.Z.

For our basic 2DPLC (with $r = 0.3a$) a frequency range exists, around $\Omega = 0.43$, where the band is almost isotropic. The equi-frequency contours, centered at the $\Gamma$ point are essentially circular in that region, and their radii decrease with increasing frequency. On that surface, an effective, isotropic refractive index can be defined $n_{eff} = -1$, and thus a negative refraction, and the unrestricted superlensisng could occur in the simple slab of the crystal [13].

In order to study the possibility of superlensing, we have performed FDTD simulations for a slab of the 2DPLC. Fig. 2 shows snapshots of field intensities produced by a point source placed to the left of the slab (at the distance $4a$), for two different orientation of the slab, horizontal $\Gamma M$ (a), and horizontal $\Gamma K$ (b). The frequency is $\Omega = 0.43$. A point image to the right of the slab forms for the horizontal $\Gamma M$ orientation of the slab (a). The absence of the image in the other direction (b) is due to the symmetry mismatch between the incident field and the symmetry allowed for mode propagation inside the crystal. Similar image suppression has been studied earlier [21].

The resolution of our lens can be estimated from the calculated time-average field intensity (shown in Fig. 3) across the image center, parallel to the 2DPLC slab of Fig. 2(a). The resolution, defined here as the full width at half maximum (FWHM), is about $0.3\lambda$, for the case of $\Gamma = 0$ (metal with no loss). This obviously is a subwavelength resolution. For $\Gamma \neq 0$, i.e. when the absorption losses are present in the metal, the image curves in Fig. 3 broaden, and the resolution is reduced, as expected. For example, the resolution of $\sim 0.5\lambda$ is still observed for a large $\Gamma = 0.04\omega_p$. We confirm that the image formation follows the rules of geometric optic with $n_{eff} = -1$, by studying the image formation with slabs of different thicknesses [13,15].

Existence of surface modes in a crystal slab can explain the subwavelength lensing, as discussed in Ref. 2. In order to demonstrate that such surface modes are indeed excited in our lenses, we investigate a narrow slab of 2DPLC $0.5\lambda$ wide and $20\lambda$ long, as shown in Fig. 4 (left panel). A point source with frequency $\Omega = 0.43$ was placed a distance $0.2\lambda$ from the left interface of the slab. Surface modes are clearly visible, with alternating extrema of the electric field periodically distributed along both surfaces. These modes are surface localized, with their field intensities decaying sharply away from the surfaces. This is clearly visible in the middle-top, 3D plot panel, and even clearer from the intensity distribution along the dotted line, shown in the middle-bottom panel. Presence of the surface modes facilitates, at least partial compensation of the decaying evanescent waves [2]. In the narrow 2DPLC of Fig. 4 this compensation is stronger than that in the 2DPLC shown in Fig. 2, and consequently the resolution is improved to $0.2\lambda$, as shown in Fig. 3 (bold solid line).

There is a remarkable flexibility in the design parameters of the 2DPLC, which allows for achieving the desired subwavelength lensing for a variety of materials. By properly choosing the rod material, rod radius, lattice constant, and the background matrix refractive index, one can achieve the lesning at any chosen frequency in the visible and infrared ranges. For example, a slab of the 2DPLC (described above) made of Al rods, immersed in a TiO₂ matrix ($\varepsilon = 5$), will superlens around $\lambda = a/0.43$, e.g. at $\lambda = 0.65\mu m$ for $a = 0.28\mu m$. Here, the dielectric function of Al is assumed to be Drude, with $\varepsilon_b \approx 1$, $\hbar\omega_p \approx 15eV$, and $\Gamma = 0.01\omega_p$. A similar range is expected for the same 2DPLC crystal, but with Al rods replaced with carbon nanotubes, or Ag. Slight adjustments to the rod radii are then necessary, to account for the different plasma frequencies, e.g. $r = 0.4a$ for Ag.

In conclusion, we show by numerically solving Maxwell equations, that in a two-dimensional polaritonic crystal a negative refraction can occur, and a simple, thin slab of this material can act a lens, capable of imaging with subwavelength resolution. We propose this 2DPLC system, made from available materials such as $TiO_2$, and Ag, Al, or carbon nanotubes, for an experimental demonstration of these phenomena in the visible frequency range.

This work is supported by US Army Research Development and Engineering Command, Natick Soldier Center, under the grant DAAD16-03-C-0052. X. W. acknowledges C. W. Niu from MIT for simulation help.

**Figure Captions:**

**Fig. 1** Band structure for TM modes of a triangular 2DPLC. The parameters are $r = 0.3a$ (solid line), $r = 0.25a$ (dashed line), $\varepsilon = 5$. Circles are for a model based on Eq. (3).

**Fig. 2** (Color online) Snapshots of field intensities produced by a point source placed to the left of the slab (at the distance $4a$), for two different orientation of the slab, horizontal $\Gamma M$ (a), and horizontal $\Gamma K$ (b). The normalized frequency is $\Omega = 0.43$.

**Fig. 3** (Color online) The average field intensity (normalized to the source intensity) across the image center, parallel to the 2DPLC slab of Fig. 2 (a) for different $\Gamma$. Thin solid line is for $\Gamma = 0$, dashed for $\Gamma = 0.01\omega_p$, and dash-dotted for $\Gamma = 0.04\omega_p$. Bold solid line is for a thinner slab, shown in Fig. 4, where well defined surface modes exist.

**Fig. 4** (Color online) The field intensity map for a narrow slab of 2DPLC $0.5\lambda$ wide, and $20\lambda$ long (left panel). The middle-top panel shows a 3D view of the part of the field map in the vicinity of the source and the image. Surface modes are clearly visible in both panels, with alternating extrema of the electric field periodically distributed along both surfaces. The average field intensity along the dotted line of the middle-top panel clearly show the sharp decay of the electric field towards the center of the slab (middle-bottom panel).

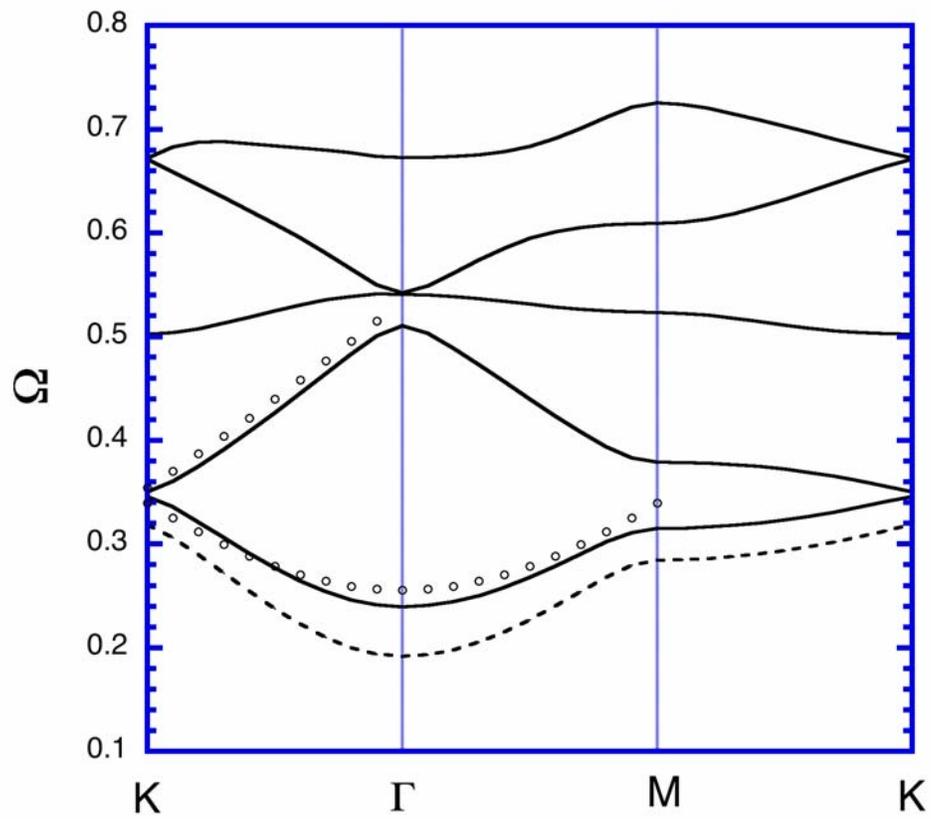

Fig. 1
X. Wang and K. Kempa

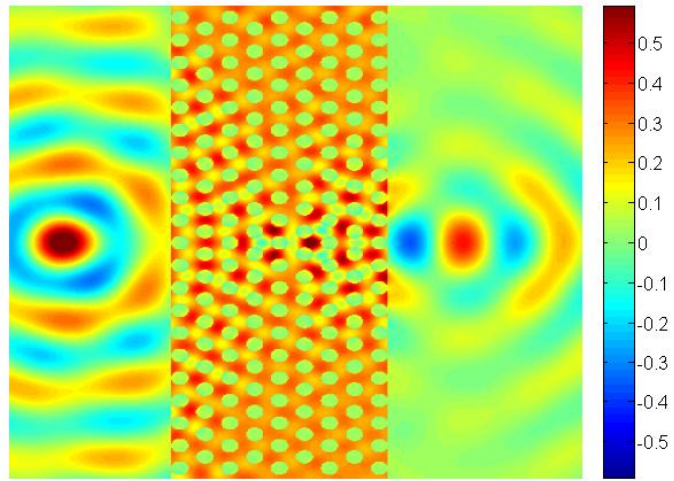

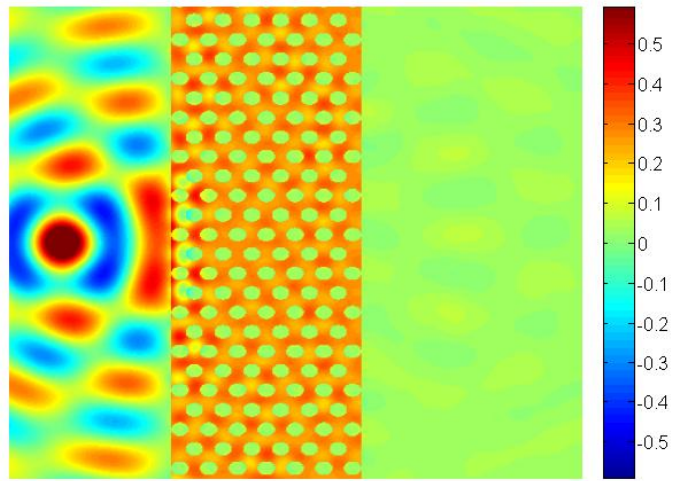

Fig. 2
X. Wang and K. Kempa

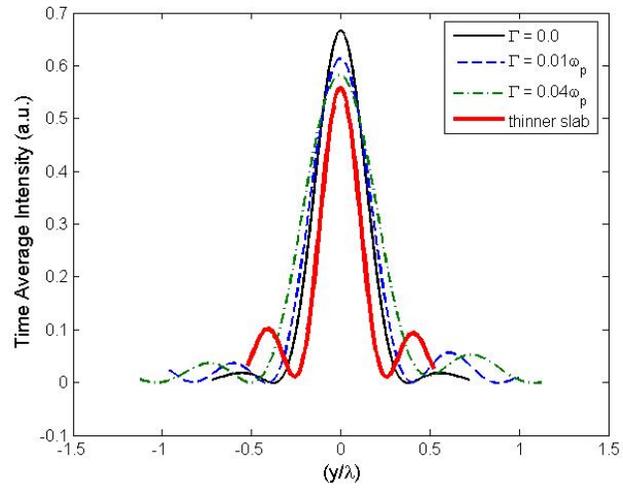

Fig. 3
X. Wang and K. Kempa

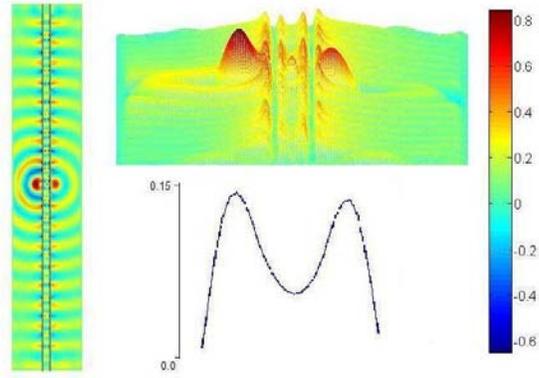

Fig. 4
X. Wang and K. Kempa